\documentstyle[12pt]{article}
\begin{document}
\thispagestyle{empty}

\def\ve#1{\mid #1\rangle}
\def\vc#1{\langle #1\mid}

\newcommand{\p}[1]{(\ref{#1})}
\newcommand{\be}{\begin{equation}}
\newcommand{\ee}{\end{equation}}
\newcommand{\sect}[1]{\setcounter{equation}{0}\section{#1}}

\renewcommand{\theequation}{\thesection.\arabic{equation}}

\newcommand{\vs}[1]{\rule[- #1 mm]{0mm}{#1 mm}}
\newcommand{\hs}[1]{\hspace{#1mm}}
\newcommand{\mb}[1]{\hs{5}\mbox{#1}\hs{5}}
\newcommand{\Db}{{\overline D}}
\newcommand{\bea}{\begin{eqnarray}}
\newcommand{\eea}{\end{eqnarray}}
\newcommand{\wt}[1]{\widetilde{#1}}
\newcommand{\und}[1]{\underline{#1}}
\newcommand{\ov}[1]{\overline{#1}}
\newcommand{\sm}[2]{\frac{\mbox{\footnotesize #1}\vs{-2}}
           {\vs{-2}\mbox{\footnotesize #2}}}
\newcommand{\prt}{\partial}
\newcommand{\eps}{\epsilon}

\newcommand{\R}{\mbox{\rule{0.2mm}{2.8mm}\hspace{-1.5mm} R}}
\newcommand{\Z}{Z\hspace{-2mm}Z}

\newcommand{\cd}{{\cal D}}
\newcommand{\cg}{{\cal G}}
\newcommand{\ck}{{\cal K}}
\newcommand{\cw}{{\cal W}}

\newcommand{\vj}{\vec{J}}
\newcommand{\vl}{\vec{\lambda}}
\newcommand{\vz}{\vec{\sigma}}
\newcommand{\vt}{\vec{\tau}}
\newcommand{\vw}{\vec{W}}
\newcommand{\poiss}{\stackrel{\otimes}{,}}

\def\l#1#2{\raisebox{.2ex}{$\displaystyle
  \mathop{#1}^{{\scriptstyle #2}\rightarrow}$}}
\def\r#1#2{\raisebox{.2ex}{$\displaystyle
 \mathop{#1}^{\leftarrow {\scriptstyle #2}}$}}

\renewcommand{\thefootnote}{\fnsymbol{footnote}}
\newpage
\setcounter{page}{0}
\pagestyle{empty}
%\begin{flushright}
%{February 1998}\\
%{Mexico ???}\\

%{JINR E2-98-???}\\
%{solv-int/0203225}
%\end{flushright}

\vs{8}

\begin{center}

{\LARGE {\bf Some integrable models in quantized spaces}}\\

\vs{8}

\large A.N. Leznov$^{1,2,3}$\\[.5em]
{\small \it $^{(1)}$Universidad Autonoma del Estado de Morelos,
CCICAp,}\\[-.5em]
{\small \it Cuernavaca, Mexico}\\
{\small \it $^{(2)}$Institute for High Energy Physics}\\[-.5em]
{\small \it 142280, Protvino, Moscow Region, Russia}\\
{\small \it $^{(3)}$Bogoliubov Laboratory of Theoretical Physics,
JINR,}\\[-.5em]
{\small \it 141980 Dubna, Moscow Region, Russia}
\end{center}

\vs{8}

\begin{abstract}

It is shown that in a quantized space determined by the $B_2\quad
(O(5)=Sp(4))$ algebra with three dimensional parameters of the length
$L^2$,
momentum $(Mc)^2$, and action $S$, the spectrum of the Coulomb problem
with
conserving Runge-Lenz vector coincides with the spectrum found by
Schr\"odinger for the space of constant curvature but with the values
of the
principal quantum number limited from the side of higher values. The
same
problem is solved for the spectrum of a harmonic oscillator.

\end{abstract}

\vfill

{\em E-Mail:\
andrey@buzon.uaem.mx}

\newpage
\pagestyle{plain}

\renewcommand{\thefootnote}{\arabic{footnote}}

\setcounter{footnote}{0}

\section{Introduction}

The idea of uniting a manifold with its invariant group
(the group of its motions) seems to be necessary found in the
Erlangent
program by F.Klein \cite{1}.

All physical phenomena take place in a four-dimensional manifold
described by four space-time coordinates $x_i$ and a 10-parametric
group of
its motions (the Poincare group) with 4 generators of the
displacements $p_i$
and 6 generators of the Lorentz rotations $F_{ij}$.

All these objects obviously satisfy the following system of
commutation
relations
$$
[p_i,x_j]=ihg_{ij}I,\quad  [p_i,p_j]=0,\quad [x_i,x_j]=0,
$$
$$
[I,x_j]=0, \quad [I,p_i]=0, \quad [I,F_{ij}]=0
$$
\be
[F_{ij},x_s]=ih(g_{is}x_j-g_{js}x_i),\quad
[F_{ij},p_s]=ih(g_{is}p_j-g_{js}p_i) \label{1}
\ee
$$
[F_{ij},F_{sk}]=ih(g_{js}F_{ik}-g_{is}F_{jk}-g_{jk}F_{is}+g_{ik}F_{js}
)
$$
In (\ref{1}) $h$, is the Planck constant, $g_{ik}$ is the signature of
the
four-dimensional coordinate space, and $I$ is the unity element of the
algebra commuting with all its elements (and so belonging to the
center of
the algebra).

Algebra (\ref{1}) consists of 15 elements, possesses 3 Casimir
operators,
and, when the spin  generators $S_{ij}=F_{ij}-x_ip_j+x_jp_i$
(the total moment minus its orbital part) are introduced, can be
represented
in the form
$$ K_1=I,\quad K_2=\sum S_{ij}S_{ij},\quad K_3=\sum
\epsilon_{ijkl}S_{ij}S_{kl} $$
Hence, it follows that irreducible
representations of algebra (\ref{1}) can be realized on the basis
functions
$\psi^{s_1,s_2}_{\alpha,\beta}(x)$.  The action of generators of the
algebra
in this basis has the following form
$$ p_k=ih{\partial\over \partial
{x_k}},\quad F_{ij}=x_ip_j-x_jp_i+S_{ij} $$
where $x_k$ act as operators of
multiplication, and operators $S_{ij}$ act on spin indices
$\alpha,\beta$ of
the wave function $\psi^{s_1,s_2}$.

The  algebra (\ref{1}) is transformed to the algebra of a quantum
space
under the following two conditions:
The dimensions and meaning of the generators (as coordinates, momenta,
and so on) should be preserved; 2. the new algebra should contain the
Lorentz
algebra as its subalgebra. So the property of the Lorentz symmetry is
conserved as a fundamental law of the nature. Implicitly, it is also
assumed
that the algebra should be a linear one. The last condition is not
necessary, but in what follows, we will assume that it takes place.

The most general form of the commutation relations of the quantum
space
is the following:
$$
[p_i,x_j]=ih(g_{ij}I+{F_{ij}\over S}),\quad  [p_i,p_j]={ih\over
L^2}F_{ij},
\quad [x_i,x_j]={ih\over M^2}F_{ij},
$$
\be
[I,p_i]=ih({p_i\over S}-{x_i\over L^2}), \quad [I,x_i]=ih({p_i\over
M^2}-
{x_i\over S}),\quad [I,F_{ij}]=0 \label{2}
\ee
$$
[F_{ij},x_s]=ih(g_{is}x_j-g_{js}x_i),\quad
[F_{ij},p_s]=ih(g_{is}p_j-g_{js}p_i)
$$
$$
[F_{ij},F_{sk}]=ih(g_{js}F_{ik}-g_{is}F_{jk}-g_{jk}F_{is}+g_{ik}F_{js}
)
$$
Commutation relations of the quantum space contain 3 dimensional
parameters:
the dimension length $L$, the momentum $Mc\to M$, and the action $S$.
The
Jacobi equalities are satisfied for (\ref{2}). If it is assumed that
the
transition to the classical dynamics is realized by the conventional
change:
$$
{1\over h} [A,B]\to \{A,B\}
$$
(transition from commutators to the Poisson brackets), then it is
necessary
to conclude that values of the dimensional parameters are on the
cosmic
scale in order to have usual dynamics at least at distances of the
order of
the Sun system.

The limiting procedure $ M^2,S\to \inf$ leads to the space of
constant curvature considered in the connection with the Coulomb
problem by
E.~Schr\"odinger \cite{2}; $L^2,S\to \inf$ leads to the quantum space
considered by Snyder \cite{3}; and $S\to \inf$ leads to the Yang
quantum
space \cite{4}.  The parameter of action $S$ additional to $L^2,M^2$
was
introduced in \cite{5}.

The term "quantum (or quantized) space" is not very useful, because
the
modified classical dynamics can be considered in it (as well as
electrodynamics, gravity theory and, so on).

\section{Quantum space in 3 dimensions}

Since in what follows, we will consider the three-dimensional
stationary problems, we would like here to examine
quantum spaces in three dimensions more accurately. In this case, the
system
(\ref{2}) is reduced to $$
[p_i,x_j]=ih(\delta_{ij}I+{\epsilon_{ijk}i_k\over
S}),\quad [p_i,p_j]= {ih\over L^2}\epsilon_{ijk}l_k,\quad
[x_i,x_j]={ih\over
M^2}\epsilon_{ijk}l_k, $$ \be [I,p_i]=ih(-{p_i\over S}+{x_i\over
L^2}), \quad
[I,x_i]=ih(-{p_i\over M^2}+ {x_i\over S}),\quad [I,l_i]=0 \label{3}
\ee $$
[l_i,x_s]=ih\epsilon_{sij}x_j,\quad [l_i,p_s]=ih\epsilon_{sij}p_j
$$
$$
[l_i,l_s]=ih\epsilon_{isj}l_j
$$
Algebra (\ref{3}) is 10-dimensional and possesses two Casimir
operators.
The quadratic Casimir operator is of the form
$$
K_2=I^2+{x^2\over L^2}+{p^2\over M^2}-{px+xp\over S}+l^2(-{1\over
S^2}+
{1\over L^2M^2})
$$
where $a^2=a_1^2+a_2^2+a_3^2$, and the fact of commutation of $K_2$
with all
generators of the algebra (\ref{3}) can be checked by a direct
calculation.

The Cartan-Killing form $K_2$ is quadratic in the generators of
the algebra and can be transformed to the main axes, for instance, in
the
following way \be K_2=I^2+x^2({1\over L^2}-{M^2\over S^2})+{1\over
M^2}(p_i-M^2{x_i\over S})^2+ l^2(-{1\over S^2}+{1\over
L^2M^2})\label{K} \ee
We draw attention of the reader that the signs of the constants
$M^2,L^2$
are fixed just as the sign of dimensionless constant $(1-({LM\over
S})^2)$
in the expression for $K_2$. Depending on its choice, there are
possible the
foliowing semicompact algebras are possible: $O(5),O(1,4),O(2,3)$ and
their
limiting cases when dimensional parameters are going to infinity, as
it was
mentioned in the Introduction. For definiteness, in what follows we
restrict
ourselves to the case of $O(5)$, which corresponds to the choice
$0\leq L^2,
0\leq M^2, 0\leq 1-({LM\over S})^2$.

To understand the meaning of the algebra (\ref{3}) more precisely, it
is
useful to carry out the following identification:
\be I=ih F_{45},\quad
x_i={ih\over M}F_{4i},\quad p_i-M^2{x_i\over S}={ih\over L}
F_{5i},\quad
l_i={ih\over 2}\epsilon (i,k,l)F_{kl}\label{B} \ee where
$F_{\alpha,\beta}$
are the generators of $B_2$ ($O(5)$) algebra.

To derive an explicit expression for the Casimir operator of the
fourth order,
let us consider the vector
$$
ihS_i=[p_i,(xl)]=[(pl),x_i],\quad S_i=\epsilon_{ijk}x_jp_k+(I-{ih\over
S})l_i=
-\epsilon_{ijk}p_jx_k+(I+{ih\over S})l_i
$$
where $(AB)=A_1B_1+A_2B_2+A_3B_3$. The equivalence of two forms for
the vector
$S$ can be checked by a direct calculation. From this it follows
that in the limit of infinite values of dimensional parameters, this
vector
passes into the spin vector of the usual theory. By direct
calculations, one
can be convinced that this vector satisfies   the following
commutation
relations $$ [S_i,x_s]=\delta_{is}[I,(xl)]=\delta_{is}(-{(pl)\over
M^2}+{(xl)\over S}),\quad
[S_i,p_s]=\delta_{is}[I,(pl)]=\delta_{is}(-{(pl)\over S}+{(xl)\over
L^2})
$$
$$
[I,S_i]=0,\quad [S_i,x_j]=[S_j,x_i],\quad [S_i,p_j]=[S_j,p_i]
$$
Using the above relations, it is not difficult to check that the
Casimir
operator of the fourth order has the form $$ K_4=\sum
S_i^2-{(pl)^2\over
M^2}+{(pl)(xl)+(xl)(pl)\over S}-{(xl)^2\over L^2} $$ From this
expression
it follows that in the scalar representation ($K_4=0$), the equalities
$(lp)=(lx)=0$ are satisfied.

\section{Scalar representation of algebra of the quantum space}

Connection between the elements of a quantized space with generators
of $O(5)$
algebra (\ref{B}) makes it possible to express them explicitly with
the help
of the well-known Gelfand-Zeitlin formulae \cite{6} for irreducible
representations of $O(5)$ algebra.  We will require only the
matrix elements of the following operators $I, px+xp, x^2,p^2,
l^2,(lp),(lx)$
invariant with respect to the rotation group. In what follows, it will
be
more suitable to find them independently. As it is mentioned above,
the
scalar representation obeys the relations $$ (lp)=(lx)=0,\quad
x_ip_j-x_jp_i=-(I-{ih\over S})\epsilon_{ijk}l_k,\quad
p_ix_j-p_jx_i=(I+{ih\over S})\epsilon_{ijk}l_k
$$
With the help of the last relations, the (nonlinear) algebra of the
commutation relations for the quantities we are interested in can be
written
as (of course, in the process of calculation it is necessary to use
(\ref{3}))
$$ [I,p^2]=ih({D\over L^2}-{2p^2\over S}),\quad
[I,x^2]=ih(-{D\over M^2}+{2x^2 \over S}),\quad [I,D]=2ih({x^2\over
L^2}-{p^2\over M^2}) $$ \be [p^2,x^2]=ih(DI+ID+4{l^2\over
S}+6{h^2\over
S}),\label{QA} \ee $$ [D,x^2]=i2h((x^2+{l^2\over
M^2})I+I(x^2+{l^2\over
M^2})+{3h^2\over 2M^2}) $$ $$ [D,p^2]=2ih((p^2+{l^2\over
L^2})I+I(p^2+{l^2\over L^2})+{3h^2\over 2L^2}) $$ where the notation
for
the dilation operator $D=px+xp$ was introduced. The nonlinear algebra
(\ref{QA}) consists of 5 elements and possesses 3 Casimir
operators-two of
the initial algebra of the quantized space (one of which is equal to
zero in
its scalar representation) and the operator $l^2$ belonging to the
center of
this algebra.  So, it can be realized on the one-dimensional space.
For it,
we choose the proper values of the operator $r^2=x^2+{l^2\over M^2}+
{h^2\over M^2}$ that differs from the usual form of the
Casimir operator of the algebra of four-dimensional rotations with the
generators $(Mx_i,l_i)$ (see (\ref{3}) by the last term. Thus, we
conclude
that $$ (r^2)_{n,n}={h^2\over M^2}(n+1)^2 $$ This formula allows us to
to speak about the matrix elements of the $r$ operator with diagonal
matrix elements $r_{n,n}={h\over M}(n+1)$ and without a singularity at
the
origin $n=0$.

The commutation relations for the $x^2$ generator (\ref{QA}) rewritten
in
terms of the operator $r^2$
$$
[I,r^2]=ih({2r^2\over S}-({D+2{l^2\over S}+2{h^2\over S})\over M^2}),
\quad[D,r^2]=ih(2(r^2I+Ir^2)-{h^2\over M^2}I)
$$
make it possible to connect the matrix elements of $I$ and $D$
operators (different from zero) and once again in a different way to
obtain
diagonal matrix elements of the $r^2$ operator
$$
D_{n+1,n}=-ih(2n+3)I_{n+1,n},\quad D_{n-1,n}=ih(2n+1)I_{n-1,n},
$$
$$
D_{n,n}={2h^2\over S}((n+1)^2-l(l+1)-1),\quad I_{n,n}=0
$$

Commutation relations (\ref{QA}) with $p^2$ give a
possibility to calculate the matrix elements of this operator via
matrix
elements of the $I$ generator
$$
p^2_{n,n}={h^2\over
L^2}((n+1)^2-l(l+1)-1)+M^2((2n+3)I_{n+1,n}I_{n,n+1}-
(2n+1)I_{n,n-1}I_{n-1,n})
$$
$$
p^2_{n-1,n}={ih^2\over S}M^2(2n+1)I_{n-1,n},\quad
p^2_{n,n-1}={ih\over S}M^2(2n+1)I_{n,n-1}
$$
$$
p^2_{n-2,n}=-M^2I_{n-2,n-1}I_{n-1,n},\quad
p^2_{n,n-2}=-M^2I_{n,n-1}I_{n-1,n-2}
$$

The quadratic Casimir operator (\ref{K}) allows us to calculate the
explicit
form for matrix elements of the generator $I$
\be
K_2=2(n+2)I_{n,n+1}I_{n+1,n}-2nI_{n,n-1}I_{n-1,n}+h^2({1\over
M^2L^2}-{1\over
S^2})(2(n+1)^2-l(l+1)-2)\label{KI}
\ee
It is necessary to keep in mind that $l\leq n$, and for this reason
the
matrix element $I_{n,n-1}(l)$ equals zero if $n=l$.  Multiplying both
sides
of the last equality by $(n+1)$ and summing up the result from
$n=l$ up to $n+1$, we obtain $$
K_2((n+1)(n+2)-l(l+1))=4(n+1)(n+2)I_{n,n+1}I_{n+1,n}+h^2({1\over
M^2L^2}-
{1\over S^2})\times
$$
$$
([(n+1)(n+2)]^2-[l(l+1)]^2-[l(l+1)+2]((n+1)(n+2)-l(l+1))
$$
In the process of calculation, we have used the following formulae of
summation
$$
\sum_{k=1}^n k={n(n+1)\over 2}\quad \sum_{k=1}^n k^3=[{n(n+1)\over
2}]^2
$$
Introducing the notation $0\leq \delta^2\equiv h^2({1\over
M^2L^2}-{1\over S^2}
)$ and representing $K_2$ as $K_2=\delta^2\sigma(\sigma+3)$, we obtain
for the matrix element $I_{n,n+1}I_{n+1,n}$
$$
4I_{n.n+1}I_{n+1,n}=\delta^2{(\sigma-n)(\sigma+n+3)(n-l+1)(n+l+2)
\over (n+1)(n+2)}
$$
Omitting the phase factor for the matrix element $I_{n,n+1}$, we
obtain
finally \be I_{n,n+1}={\delta\over 2}
\sqrt{{(\sigma-n)(\sigma+n+3)(n-l+1)(n+l+2)\over
(n+1)(n+2)}}\label{OI} \ee
In terms of matrix elements of this operator, the matrix elements of
all
generators of the algebra of the quantized space can be expressed via
the
formulae of this section.  In the last formula and before, $\sigma=N$
is a
positive natural number.  Only with this choice of $\sigma$, the
constructed
representation is unitary, and all the generators involved would be
the
Hermitian ones.

\section{The guess of the Runge-Lenz vector}

A priori, we do not know the explicit form of the Coulomb potential
in the case of the quantized space, and there is an infinite number of
possibilities to write down its generalization with a correct limit
after
passing to the usual space. But we assume that the symmetry properties
of a dynamic system are the most fundamental criterion and that
these properties must be conserved in the case of the quantized space.
We
know that the usual Kepler problem is invariant with respect to the
Runge-Lenz symmetry described by the vector of the same name.

So, we begin our investigation with the explicit form of the
Runge-Lenz vector
$A_i$ in the case of the quantum space, taking into account the
additional
noncommutativity of its components connected with coordinates ($x_i$).
Below, we present the explicit form for components of the Runge-Lenz
vector
(see, for instance, \cite{LL})
$$
A_1=l_2p_3-p_2l_3+{\alpha x_1\over r}\equiv p_3l_2-l_3p_2+{\alpha
x_1\over r}
$$
\be
A_2=l_3p_1-p_3l_1+{\alpha x_2\over r}\equiv p_1l_3-l_1p_3+{\alpha
x_2\over r}
\label{RL}
\ee
$$
A_3=l_1p_2-p_1l_2+{\alpha x_3\over r}\equiv p_2l_1-l_1p_2+{\alpha
x_3\over r}
$$
Using the above definition, it is not difficult to calculate the
commutation
relations between components of the R-L vector in the scalar
representation
of the quantum space ($(xl)=(pl)=0,x_ip_j-x_jp_i=-(I-{ih\over
S})\epsilon_
{ijk}l_k$) with the result
\be
[A_i,A_j]=ih({\alpha^2\over M^2r^2}-{i\over h}({\alpha \over
r}(xp)-(px)
{\alpha \over r})-p^2+{l^2\over L^2})\epsilon_{ijs}l_s\label{QRL}
\ee
By the same technique  for the square  of the R-L vector, we derive
\be
A^2={\alpha^2x^2\over r^2}+ih({\alpha \over r}(xp)-(px){\alpha \over
r})
+h^2(p^2-{l^2\over L^2})+l^2({\alpha \over r}I+I{\alpha \over
r})+l^2p^2
\label{A}
\ee
Using the explicit form of the matrix elements of quantized space from
the
previous section, we come to the following fundamental (for us)
equality
$$
{i\over h}({\alpha \over r}(xp)-(px){\alpha \over r})={\alpha \over
r}I+
I{\alpha\over r}
$$
Taking into acount the definition of $r^2=x^2+{l^2\over M^2}+{h^2\over
M^2}$
and the last equality, we can rewrite the square of the Runge-Lenz
vector as
follows
\be
A^2=\alpha^2+2E(l^2+h^2)-{l^2\over L^2}(l^2+2h^2)\label{RLI}
\ee
where
\be
2E=(p^2+{l^2\over L^2}+({\alpha \over r}I+I{\alpha \over r})-{\alpha^2
\over r^2M^2})\label{SCH}
\ee
Now we would like to show that each component of the R-L vector
commutes with
the $E$ operator. With the help of the known matrix elements of the
algebra
of quantum space of the previous section, it is possible to check
this directly.  A shorter way consists in the consideration of the
commutator
$[A^2,A_i]$. Thus, using (\ref{RLI}), we have for both the sides of
this equality $$ [A^2,A_j]=A_i[A_i,A_j]+[A_i,A_j]A_i=2ih([A\times
l]_j(E-{l^2\over l^2})- 2ih(E-{l^2\over l^2})[l\times A]_j= $$ $$
2[E,A_j](h^2+l^2)+2E([A\times l]_j-[l\times A]_j) -([A\times
l]_j-[l\times A]
_j)){l^2+h^2\over L^2}-{l^2\over L^2}([A\times l]_j-[l\times A]_j)
$$
After equating similar terms from both the sides of the last equality,
we arrive at the equation for the vector $[E,A_i]\equiv B_i$
$$
B_j(l^2+h^2)+ih[B\times l]_j=0,
$$
>from which it immediately follows that
$$
(Bl)=0
$$
For the vector product of the vector $B$ by $l$, we find
$$
[B\times l]=iB
$$ or
$B=[E,A]=0$, or in other words, the operator $E$ commutes with every
component
of the R-L vector. This fact allows us to use the technique of the
paper
\cite{7} (the reader can there find the  corresponding references on
the
same subject) and to obtain the spectrum of the energy operator $E$.

Thus, in the case of the quantum space with dimensional parameters
$M,L,S$,
the spectrum of this operator depends only on the $L$ parameter and
has the
form found by Schr\"odinger in ref. [?] for the space of constant
curvature
\be 2E_n=-{\alpha^2\over h^2n^2}+{h^2\over L^2}(n^2-1),\quad l\leq n
\label{RID}
\ee

But in the case of the quantum space of the general position, the
values
of $n$ are restricted by the condition
$$
l\leq n \leq N
$$
that will be clear from the results of the next section.

We would like to notice also that in the generalized Snyder quantum
space
($L^2\to \inf$), the spectrum of the hydrogen atom exactly coincides
with the
spectrum in the case of the flat space. And so, in principle, by the
spectroscopic data, these two possibilities cannot be distinguished.

\section{The Schr\"odinger equation of the quantum space}

The Schr\"odinger equation(\ref{SCH}) can be rewritten as
the equation in finite differences
$$
2E_n\psi_n=-M^2I_{n,n+1}I_{n+1,n+2}\psi_{n+2}+M^2({\alpha\over
hM(n+2)(n+1)}+
{ih\over S})(2n+3)I_{n,n+1}\psi_{n+1}+
$$
\be
({h^2\over
L^2}n(n+2)+M^2((2n+3)I_{n+1,n}I_{n,n+1}-(2n+1)I_{n,n-1}I_{n-1,n}-
{\alpha^2\over h^2(n+1)^2})\psi_n+\label{SCHI}
\ee
$$
M^2({\alpha\over hM(n(n+1)}-{ih\over S})(2n+1)I_{n,n-1}\psi_{n-1}-
M^2I_{n,n-1}I_{n-1,n-2}\psi_{n-2}
$$
Now we would like to find the way to solve it directly. For this aim,
let us
once more rewrite (\ref{SCHI}), introducing the abbreviation:
$\alpha\to {\alpha\over Mh}, \quad E\to {E\over M^2},\quad
e_n=-{\alpha^2\over (n+1)^2}+{h^2\over M^2 L^2}((n+1)^2-1)$
$$
2E\psi_n=-I_{n,n+1}I_{n+1,n+2}\psi_{n+2}+({\alpha\over (n+2)(n+1)}+
{ih\over S})(2n+3)I_{n,n+1}\psi_{n+1}+
$$
\be
(e_n + (2n+3)I_{n+1,n}I_{n,n+1} - (2n+1)I_{n,n-1}I_{n-1,n}) \psi_n
\label{SCHII}
\ee
$$
({\alpha\over n(n+1)}-{ih\over S})(2n+1)I_{n,n-1}\psi_{n-1}-
I_{n,n-1}I_{n-1,n-2}\psi_{n-2}
$$
Comparing this with (\ref{RID}), we conclude that to have a correct
spectrum,
the energy operator should be expressed in the form
$$
2E=UeU^{-1},\quad U^{-1}=U^*
$$
where $e$ is a diagonal matrix with diagonal elements $e_n, l\leq n
\leq N$.

The most simple way to check this assumption is to calculate
the traces of consequent degrees of the energy matrix. Using the
main relation of section 4, rewritten below (only for convenience with
the new abbreviation)
\be
I_{n.n+1}I_{n+1,n}={\delta^2\over
4}{(\sigma-n)(\sigma+n+3)(n-l+1)(n+l+2)
\over (n+1)(n+2)}\equiv x_{n+1}\label{TR}
\ee
we immediately conclude that
$$
Trace \quad 2E=\sum_{n=l}^{n=N} e_n
$$
Using the notation (\ref{TR}) and definition of the energy matrix
(\ref{SCHII}), we obtain for the trace
$$
Trace \quad (2E)^2=\sum_{n=l}^{n=N} (e_n^2+x_nx_{n-1}+x_{n+1}x_{n+2}+
$$
$$
({\alpha^2\over n^2(n+1)^2}+{h^2\over S^2})(2n+1)^2x_n+
({\alpha^2\over (n+2)^2(n+1)^2}+{h^2\over S^2})(2n+3)^2x_{n+1}+
$$
$$
2e_n((2n+3)x_{n+1}-(2n+1)x_n)+(2n+3)^2x^2_{n+1}+(2n+1)^2x^2_n-2(2n+3)(
2n+1)
x_{n+1}x_n)
$$
With all the above definitions, it is not difficult to get
$$
Trace\quad (2E)^2=\sum_{n=l}^{n=N} e_n^2
$$
We restrict ourselves to this consideration keeping in mind that the
proof of
the general formula
$$
Trace\quad (2E)^s=\sum_{n=l}^{n=N} e_n^s
$$
for arbitrary s is a good exercise for students.

\section{Harmonic oscillator}

As in the previous section, the symmetry arguments will play a major
part
in finding an explicit expression for the Hamiltonian of a harmonic
oscillator.  In the case of the usual space, this Hamiltonian commutes
with the components of the  tensor of second rank (more precisely,
with
every product containing an equal number of annihilation and creation
operators and their linear combination) $$ a_ia^*_j,\quad 1\leq i,j
\leq 3
\quad [a_i,a_j]=0, \quad [a^*_i,a^*_j]=0,\quad
[a^*_i,a_j]=\delta_{i,j} $$ $$
H={1\over 2}\sum (a_ia^*_i+a^*_ia_i)
$$
constructed of the creation and annihilation operators.

To find the generalization of this symmetry to the quantum space,
let us consider algebra (\ref{3}) introducing a linear combination of
momentum and coordinate variables (we continue our consideration just
for the case of $O(5)$ algebra of the quantum space)
$$
a_i=\delta^{-1}(p_i+\theta x_i),\quad  a^*_i=\delta^{-1}(p_i+\theta^*
x_i)
$$
Taking the parameter $\theta$ to satisfy the conditions
$$
[a_i,a_j]=0, \quad [a^*_i,a^*_j]=0
$$
(the factor $\delta$ is introduced only with the aim not to rewrite
the same
formulae once more).  This is equivalent to the following quadratic
equation $$ {\theta^2\over M^2}+{2\theta\over S}+{1\over L^2}=0,\quad
{(\theta^*)^2\over M^2}+{2\theta^*\over S}+{1\over L^2}=0
$$
with the solution
$$
{\theta\over M^2}=-{1\over S}+{i\delta\over h}
$$
It is necessary to emphasize that the operators $a_i,a^*_i$ belong to
the
algebra $O(5,C)$, rather than to $O(5,C)$, but this is not essential
in what
follows.  By a simple calculation, it is possible to check that the
constructed operators $a_i,a^*_i$ satisfy the commutation relations $$
[I,a_i]=-\delta a_i,\quad [I,a^*_i]=\delta a^*_,\quad i[a_i,a^*_i]=-ih
(\delta_{i,j}\hat I+{\epsilon(i,j,k)l_k\over S} $$ where  $I=\delta
\hat I$.
In the new notation, the Casimir operator of second order (\ref{K})
takes the form $$ K_2=I^2+\delta^2
M^2(\sum(a_ia^*_i+a^*_ia_i)+\delta^2{l^2\over h^2}\equiv I^2+H $$

>From the last expression, it follows that the Hamiltonian $H$
commutes
with all polynomials constructed of creation and annihilation
operators
containing an equal number of them.

The eigenvalues are determined from the equation
$$
H\psi=(N(N+3)-\hat I^2)\psi
$$
The eigenvalues of the operator $I^2$ are equal to the squares of
natural numbers $m^2$ limited by the condition
$$
m^2\leq (N-l)^2
$$
In the classical limit (the usual three-dimensional oscillator), the
spectrum
changes to the well-known spectrum with a linear dependence on the
principal
quantum number.

\section{Outlook}

The main result of the present paper consists in finding the rules for
working in quantized spaces with the dimensional parameters
$L^2,M^2,S$.
We have shown that, in the general case of a quantized space, the
energy spectrum of the hydrogen atom (the Kepler problem) is
exactly the same as that obtained by Schr\"odinger in his known paper
\cite{2} who has considered this problem in the space of constant
curvature more than 50 years ago. The only difference consists in
that this spectrum is limited from the side of large principal quantum
numbers of the hydrogen atom.

It seems interesting to supplement the consideration with the
interaction
with a monopole field (the so-called MIC problem), like it has been
carried out in \cite{7}, and it is obvious that this is possible
without any big difficulties. It is also interesting to consider all
other
integrable models in quantized spaces and to clearly understand the
general
connection between these two background foundations of the theory of
quantization of the space and integrability.

We pay attention of the reader that the algebra of the quantized space
was
chosen in the form (\ref{3}) under an implicit assumption of its
linearity.
But it is not less interesting to assume that all coefficient
functions of
the algebra are nonlinear and depend on Lorentz-invariant quantities
such as
$p^2,x^2,..$ and all other possible Lorentz-invariant combinations.

The next obvious generalization consists in the possibility to
include, into
the game, superintegrable variables. This can be realized, for
instance, by
replacing the Poincare group of motion of the space by its supergroup
analog
\cite{8}.

It is obvious that the most interesting is the consideration and
construction
of the gauge (or supergauge) field theory on the base of the quantized
space.

And the last comment: As we have mentioned in the Introduction, the
classical
limit of the theory of quantized spaces will give a
dynamic picture of the world at distances ( and other dimensional
parameters) greater than $L^2,M^2,S$ that is absolutely different from
that predicted by the classical celestial mechanics. So, the
investigation of
this question remains a very interesting problem unsolved in the
present
paper.

\section{Acknowledgments}

The author is indebted to G.S.Pogosyan, who  paid attention of the
author
to the problem of the hydrogen atom in the space of constant curvature
with the citation of the corresponding literature and fruitful
discussions
in the process of writing the present paper.

The author is indebted to CYNNECIT for partial financial support.

\end{document}